\documentclass[sigconf]{acmart}

\definecolor{swotS}{RGB}{226,237,143}
\definecolor{swotW}{RGB}{247,193,139}
\definecolor{swotO}{RGB}{173,208,187}
\definecolor{swotT}{RGB}{192,165,184}
\usepackage{subfig}
\AtBeginDocument{%
  \providecommand\BibTeX{{%
    \normalfont B\kern-0.5em{\scshape i\kern-0.25em b}\kern-0.8em\TeX}}}

\copyrightyear{2023}
\acmYear{2023}
\setcopyright{rightsretained}
\acmConference[CHI EA '23]{Extended Abstracts of the 2023 CHI Conference on Human Factors in Computing Systems}{April 23--28, 2023}{Hamburg, Germany}
\acmBooktitle{Extended Abstracts of the 2023 CHI Conference on Human Factors in Computing Systems (CHI EA '23), April 23--28, 2023, Hamburg, Germany}\acmDOI{10.1145/3544549.3573858}
\acmISBN{978-1-4503-9422-2/23/04}

\begin{document}

\title{How to Organise Engaging Online Conferences and Escape the Zoom Rectangle}

\author{Jarosław Kowalski}
\email{jaroslaw.kowalski@opi.org.pl}
\orcid{0000-0002-1127-2832}
\affiliation{%
  \institution{LIT, National Information Processing Institute}
  \streetaddress{al. Niepodległości 188b}
  \city{Warsaw}
  \country{Poland}
  \postcode{00-608}
}

\author{Kinga Skorupska}
\email{kinga.skorupska@pja.edu.pl}
\orcid{0000-0002-9005-0348}
\affiliation{%
  \institution{XR Lab, Polish-Japanese Academy of Information Technology}
  \streetaddress{ul. Koszykowa 86}
  \city{Warsaw}
  \country{Poland}
  \postcode{02-008}
}

\author{Agata Kopacz}
\email{agata.kopacz@opi.org.pl}
\orcid{0000-0001-9337-0809}
\affiliation{%
  \institution{LIT, National Information Processing Institute}
  \streetaddress{al. Niepodległości 188b}
  \city{Warsaw}
  \country{Poland}
  \postcode{00-608}
}

\author{Bartosz Muczyński}
\email{b.muczynski@pm.szczecin.pl}
\orcid{0000-0002-0559-4181}
\affiliation{%
  \institution{Faculty of Navigation, Maritime University of Szczecin}
  \streetaddress{ul. Wały Chrobrego 1--2}
  \city{Szczecin}
  \country{Poland}
  \postcode{70-500}
}

\author{Wiesław Kopeć}
\email{wieslaw.kopec@pja.edu.pl}
\orcid{0000-0001-9132-4171}
\affiliation{%
  \institution{XR Lab, Polish-Japanese Academy of Information Technology}
  \streetaddress{ul. Koszykowa 86}
  \city{Warsaw}
  \country{Poland}
  \postcode{02-008}
}

\author{Zbigniew Bohdanowicz}
\email{zbigniew.bohdanowicz@opi.org.pl}
\orcid{0000-0002-5430-0485}
\affiliation{%
  \institution{LIT, National Information Processing Institute}
  \streetaddress{al. Niepodległości 188b}
  \city{Warsaw}
  \country{Poland}
}

\author{Gabriela Górska}
\email{gabriela.gorska@opi.org.pl}
\orcid{0000-0002-4340-5425}
\affiliation{%
  \institution{LIT, National Information Processing Institute}
  \streetaddress{al. Niepodległości 188b}
  \city{Warsaw}
  \country{Poland}
  \postcode{00-608}
}

\author{Cezary Biele}
\email{cezary.biele@opi.org.pl}
\orcid{0000-0003-4658-5510}
\affiliation{%
  \institution{LIT, National Information Processing Institute}
  \streetaddress{al. Niepodległości 188b}
  \city{Warsaw}
  \country{Poland}
  \postcode{00-608}
}


\renewcommand{\shortauthors}{Kowalski et al.}

\begin{abstract}
 As an increasing number of academic conferences transition to the online sphere, new event paradigms must be explored and developed to better utilise the unique multimedia opportunities offered by the virtual world. With this in mind, we conducted in-depth interviews with researchers, performed a SWOT analysis of remote conferences, and developed experimental conference functionalities. We implemented these during the 9th edition of a two-day international scientific IT conference, which was attended by over 277 participants on the first day and 199 on the second. In this article, we describe how these innovative functionalities met the participants' needs based on qualitative and quantitative data. We present how the experiences of remote and in-person events differ, and offer recommendations on organising remote conferences that encourage participants to exchange knowledge and engage in activities.
\end{abstract}



\begin{teaserfigure}
\centering
  \includegraphics[width=0.8\textwidth]{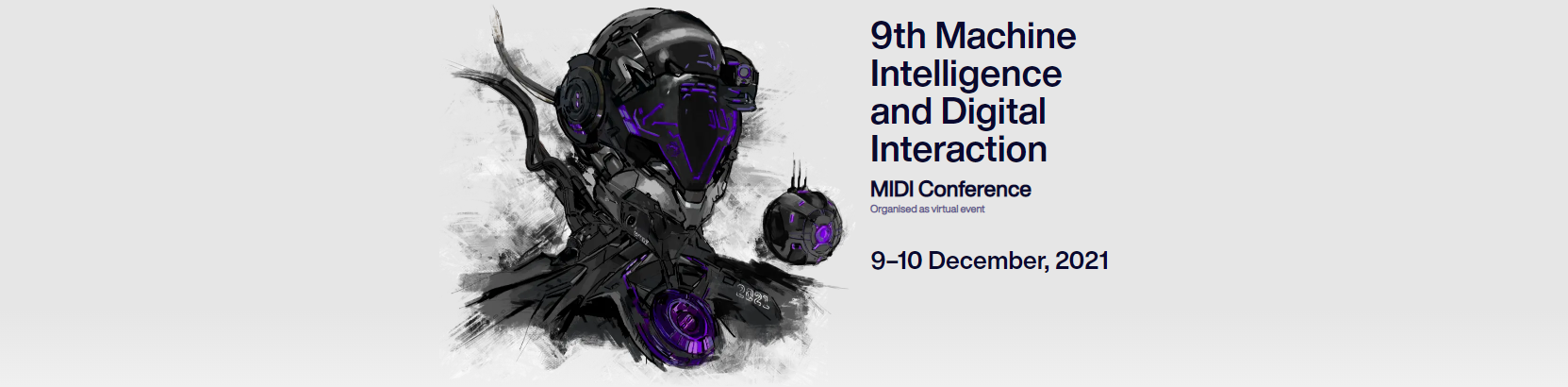}
  \caption{Our Case Study: MIDI 2021 Conference banner}
  \Description{Teaser figure presents logo and name of the 9th Machine Intelligence and Digital Interaction conference that was the focus point of the presented study.}
  \label{fig:teaser}
\end{teaserfigure}

\maketitle

\section{Rationale}
In the way remote conferences are organised at present, we can to some extent, observe skeuomorphism similar, for example, to that of the early days of the World Wide Web, when websites mimicked the layouts of more familiar media sources, such as newspapers. Websites have since taken on a life of their own, adjusting to the opportunities offered by the new medium. 
As for remote conferences, despite decades of potential virtual experiments  \cite{OnlineConfExperiment2000,onlineconferenceopensource2005,onlineconferencecase2009}, they predominantly imitate the habits of their in-person equivalents, including week-long events, thematic sessions, meeting rooms, posters, and badges. Yet, as more events transition to the online sphere \cite{virtualconferences2019,casestudyconference2021}---for reasons of health, climate, and the economy \cite{Alt-CHiCoferencingEnvironment2019,onlineconferencesinclusivity2021,injusticesonlineconfonline2021}---the necessity emerges for organisers to discover new ways of engaging audiences \cite{saliba_getting_2020,ROOS2020112975,woolston_learning_2020,pacchioni_virtual_2020}. As the primary purposes of online scientific conferences are knowledge exchange \cite{CHI2005Valueofconferences} and networking \cite{OnlineMingling2021CHI}, audience disengagement can become a major issue. The aims of this study are to explore online events more deeply and to offer guidance on the provision of remote experiences that are as engaging as in-person events by utilising the new possibilities offered by the virtual world. 

There are several proposals on how to organise virtual conferences but this mode of co-engaging with multimedia content is riddled with challenges. One of them is the frequent lack of synchronous communication between participants and one-way communication \cite{cavadas2010}. However, there is both technological and social potential for online conferences to foster interaction among participants, taking advantage of the many possibilities offered by interactive multimedia content and other ICT solutions \cite{CarrLudvi2017dt}. 

With those, and more considerations in mind, we conducted interviews, a workshop, a brainstorming session and performed a SWOT analysis of remote conferences and mapped the features of in-person conferences that increase participants' engagement. In the course of our research we selected six functionalities to introduce at the 9th edition of the Machine Intelligence and Digital Interaction (MIDI) conference, co-organised by members of our team. In this article, we offer insights into how these functionalities were used and received by the attendees. A perfect solution does not exist, and critical to the success of a remotely organised event is flexibility in implementing certain principles. In our work we offer a few specific and novel functionalities implemented at our conference and then verify their effectiveness in increasing participant engagement. We then offer recommendations and future work directions for the organisation of more engaging remote conferences.




\begin{figure*}[h!]
  \includegraphics[width=1\textwidth]{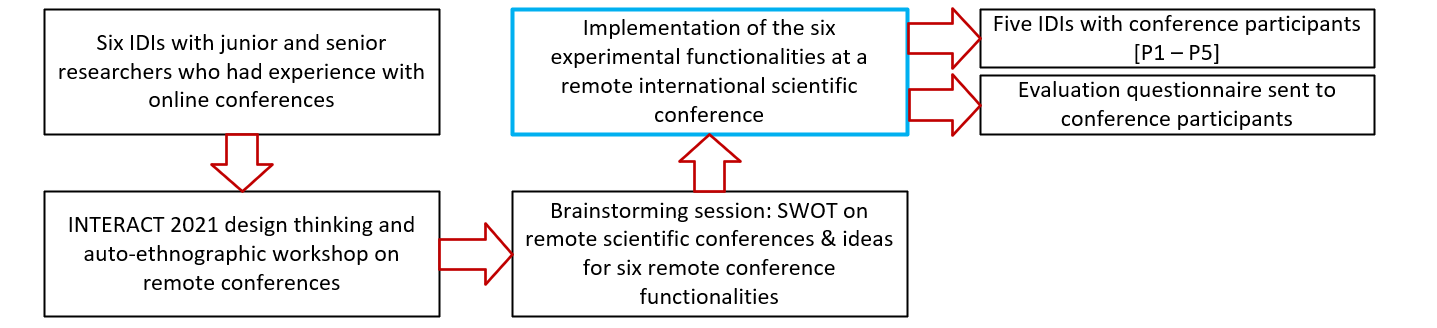}
  \caption{The research design flow}
  \Description{Figure 2 shows stages of the study design that was use by authors. In the first stage six in-depth interviews with junior and senior researchers who had experience with online conferences were conducted. In the next stage a design thinking and auto-ethnographic workshop on remote conferences was held as a part of the 18th International Conference promoted by the IFIP Technical Committee 13 on Human–Computer Interaction - INTERACT. Third stage was the brainstorming session where SWOT analysis on remote scientific conferences & ideas for six remote conference functionalities was performed. Forth stage was the implementation of the six experimental functionalities at a remote international scientific conference co-organized by authors. Two next steps were concerned with conducting Five in-depth interviews with conference participants and sending an evaluation questionnaire to conference participants.}
  \label{fig:scheme}
    \vskip -1em
\end{figure*}

\section{Methods (as depicted in Fig.\ref{fig:scheme})}

The research process commenced with six exploratory in-depth interviews (IDIs) conducted in July 2021 which involved junior and senior academics (PhD candidates, assistant professors and professors) who had experience with online conferences. The data from these ignited discussions during the following full-day workshop on the future of online conferences, which was held in August 2021. The workshop accompanied the 18th IFIP International Conference on Human-Computer Interaction. It employed design thinking methods combined with auto-ethnographic approach, which allowed participants to gather insights related to strengths and weaknesses of online conferences and to begin addressing them by proposing a preliminary set of actionable guidelines for future remote conference organization.

\subsection{Online Conferences: SWOT Analysis}

\footnotesize
\setlength{\fboxsep}{10pt}
\setlength{\fboxrule}{2pt}
\fcolorbox{gray!60}{swotS!20}{%
    \parbox{0.9\columnwidth}{%

STRENGTHS
\begin{enumerate}

\item lower costs: little or no registration fees; no travel, hotel, nor transport costs
\item potentially worldwide access---participants can easily attend conferences organised by prestigious institutions on other continents
\item better opportunities to participate without the stress of presenting
\item recorded presentations---fewer technical problems and less time pressure; no delays and recorded content can be revisited
\item lower environmental impact
\item time-saving---participation can be combined with everyday duties
\item less burdensome experience with no travel, tiredness, nor jet lag.
\end{enumerate}
        }
    }
\setlength{\fboxsep}{10pt}
\setlength{\fboxrule}{2pt}
\fcolorbox{gray!60}{swotW!20}{%
    \parbox{0.9\columnwidth}{%


WEAKNESSES
\begin{enumerate}
\item prerecorded presentations may be less engaging
\item presenters lack training on how to record engaging presentations
\item participant--participant contact is more limited
\item participants from different time zones have different activity patterns
\item fewer opportunities to make spontaneous plans with other participants
\item participants experience conferences in the same space as other activities
\item monotony and lack of variety of stimuli
\item largely passive experience, without much interaction
\item Zoom fatigue
\item individual, rather than collective experience
\item little or no opportunity for presenters to observe audiences reactions.
\end{enumerate}
        }
    }
\setlength{\fboxsep}{10pt}
\setlength{\fboxrule}{2pt}
\fcolorbox{gray!60}{swotO!20}{%
    \parbox{0.9\columnwidth}{%

OPPORTUNITIES
\begin{enumerate}
\item the opportunity to attract large numbers of authors from countries hitherto unrepresented at conferences
\item democratisation and lower barriers to entry; this is especially relevant among early-career researchers who lack access to resources (grants; funding for travel, accommodation, and registration fees) to share their research
\item lower opportunity costs of participation; ability to take part despite other responsibilities and engagements.
\end{enumerate}
        }
    }
\noindent
\setlength{\fboxsep}{10pt}
\setlength{\fboxrule}{2pt}
\fcolorbox{gray!60}{swotT!20}{%
    \parbox{0.9\columnwidth}{%
        
THREATS
\begin{enumerate}
\item difficulty maintaining focus and multitasking during conferences (participants treating conferences as streams or secondary activities)
\item communication between organisers and participants could go unnoticed
\item there may be technical problems on many ends, e.g. outages and lag
\item lack of preparation for conferences---for example, an email containing a link to join a conference and all of the relevant information, which can be read three minutes before the conference begins
\item traditional methods of communicating conference schedules are unsuitable for informing wider audiences of additional online events
\item decrease in research quality due to potentially higher limits of presenters and participants, in addition to higher acceptance rates.
\end{enumerate}
}
}

\vspace{5mm}
\normalsize

In the next step, during a brainstorming session next week, we devised functionalities that may address the weaknesses of remote conferences; the results are presented in the Strengths, Weaknesses, Opportunities and Threats (SWOT) analysis. Ultimately, six functionalities were targeted for implementation that, in part, pertain to the opportunities of remote conferencing. 
The functionalities aimed to create a situation in which participants felt that they were members of a community at the conference, reduce the feeling of anonymity, increase the chance to meet new people, and establish research collaborations.
All of the functionalities were introduced at a scientific online conference in December 2021, which was attended by 277 unique participants on the first day and 199 on the second. After the conference, participants received an evaluation questionnaire in which they were asked to rate the functionalities. We also conducted five post-conference in-depth interviews with participants who volunteered to share their impressions of the proposed functionalities.

\subsection{Case Study Design: The Conference and the Implemented Multimedia Functionalities}

The 9th edition of our international IT conference lasted two days. It was the second edition organised remotely. Based on previous experiences gathered both during 8th edition of MIDI conference and during other scientific conferences that were organized remotely, we have decided to use Zoom as one of the most common, reliable and easy to use software. The following functionalities were implemented experimentally as a part of the conference:

\subsubsection{Breakout Rooms} Following each session, separate breakout rooms were activated for ten minutes for each paper presented. In the breakout rooms, participants could converse freely with and pose questions to authors. The build-in Zoom breakout rooms feature was used. 

\subsubsection{Virtual background} Conference participants were granted the opportunity to create their own customised backgrounds using an automatic generator on the conference website. After entering their names, surnames, countries, institutions, and research interests and pressing the 'Generate' button, participants could download a PNG file containing the graphic and their information. The virtual background could be used on Zoom, which was the primary platform of the conference. This functionality was developed and implemented by conference organizers on the conference's website. 

\subsubsection{Attendee Map Pins} A functionality presented on the conference website that allowed participants to create their own 'pins' and place them on a world map. On each pin, participants could write their names, institutions, and research interests. The pins were visible to all who opened the map. After clicking on a pin, that participant's information was displayed. It was implemented by conference organizers on the conference's website using OpenStreetMap service. 

\subsubsection{Virtual Mural}
At the beginning of each conference session, the technical officer shared a link to a collaborative digital whiteboard, in which participants were encouraged to share photographs of themselves. In this way, a virtual mural was constructed, which depicted the conference attendees. A Google Jamboard virtual whiteboard was used for this purpose.

\subsubsection{Introductory Video with Tips} A short YouTube video containing advice on how to prepare for the conference and how to get the most out of the event. A link to the video was distributed in an email alongside other information about the conference. 

\subsubsection{Social Event: Geogame}
During one break, participants were invited to guess which cities were depicted in a series of photographs. The winner was the participant who successfully guessed the highest number of cities. The prize was a set of souvenirs from the conference organisers. The geogame was prepared and conducted using paid version of the Mentimeter online service which proved to be very responsive and easy to use.

\section{Results and Discussion}

\subsection{The Conference in Numbers}

\textbf{The conference took place across a Thursday and a Friday in December 2021; twenty-nine papers were presented over six sessions. The registration form remained open until the end of the conference. A total of 408 participants registered for the event; 277 attended on the first day and 199 on the second. }

A questionnaire completed by participants after the conference asked to what extent the additional functionalities contributed to four distinct objectives: 1) 'feeling that they were members of the conference community'; 2) 'getting to know new people'; 3) 'establishing research collaborations'; and 4) 'becoming better recognised by the scientific community'. The number of respondents varies because they exclusively considered the functionalities they had used or in which they had participated. These were evaluated on a scale of 1 (definitely does not contribute) to 5 (definitely contributes). Table 1  presents the mean ratings for particular functionalities and objectives. The higher the mean, the more functionality contributed to the realisation of a given objective, in the opinion of the respondents. Along with the quantitative data, we present the qualitative data with sample comments and a brief discussion of the results.


\begin{table*}[]
  \caption{Mean of answers (on a scale of 1-5) on the extent to which the activities listed in the row headers allowed the participants to achieve the effects listed in the column headers}
\begin{tabular}{p{4cm}|p{1.5cm}|p{2cm}|p{1.5cm}|p{1.5cm}|p{2cm}|}
                               & respondents      & feel that you are a member of a conference community & get to know new people & establish research collaboration & become better recognised by the scientific community \\ \hline
breakout rooms                 & n=20 & 4,25                                                 & 4,15            & 4,05                             & 4,15                                                   \\ \hline
virtual background             & n=24 & 4,17                                                 & 3,54            & 3,29                             & 3,54                                                   \\ \hline
pin of the localisation on map & n=26 & 4,04                                                 & 3,46            & 3,31                             & 3,38                                                   \\ \hline
virtual mural                  & n=17 & 4,06                                                 & 3,24            & 2,29                             & 2,53                            
                       \\ \hline
introductory movie with tips   & n=26 & 3,52                                                 & 3,08            & 2,92                             & 3,08                                                      \\ \hline  
social event - Geogame         & n=15 & 3,53                                                 & 2,80            & 2,47                             & 2,53                            
\end{tabular}

  \Description{To what extent have the activities listed below allowed you to:
(MEAN ON SCALE 1-5)}
  \label{fig:funt}
\end{table*}
\normalsize


\subsection{\textbf{Breakout Rooms}}
Although the results suggest that all of the surveyed functionalities contributed to the feeling that their users were part of a larger community of conference attendees, this was most evident in respondents' participation in breakout rooms (mean: 4.25). 
This idea was evaluated very positively. Interestingly, participants described the experience as physically 'getting closer' and 'approaching' the speakers: '\textit{Fantastic breakout rooms in which you can approach and talk---very good. I was surprised that I didn't feel restricted; that if I wanted to talk to the speaker, I could; that I could ask questions and get answers during the session}' [P3].
The interviewees' statements demonstrate that in addition to the mere existence of a space in which to talk, it is also imperative that an atmosphere be cultivated that encourages people to participate actively in discussions.
'\textit{I considered contacting others. I felt that not many people talked to each other because they were unsure of the correct etiquette. And I sometimes had this problem. [...] I was unsure what to do and of whether I would be encouraged. I think I could have been more social}' [P5].
The form of the breakout rooms involved a time limit, which was occasionally restrictive. Following expiry of the allotted time, participants in the breakout rooms were automatically transferred to the main thread of the conference; this thread was often inactive, due to the breaks between sessions. This meant that attendees who wished to skip the break and converse behind the scenes did not have the opportunity to do so. '\textit{It was cutting off, and cutting off before breaks. It was impossible to stay and talk; to drop out of the next session, if somebody wanted to. In the offline format, you can opt out of the next session---you can choose not to follow}' [P3].
In theory, a participant could make contact with others outside of the conference platform (e.g. via email); unless, however, they had exchanged email addresses during conversation in the breakout room, this entailed some difficulty. Social media could be utilised in the same way. One respondent indicated that a session chair had been asked for the email address of an author, but possessed no such information [P1].
The user experiences of networking described above suggest that post-session breakout rooms alone---which, although warmly welcomed, are limited to ten minutes---are unable to adequately meet all users' networking needs.

\subsection{\textbf{Virtual Background---generated 92 times}}
Alongside the breakout room feature, use of the virtual background contributed the most to participants' sense of belonging to a larger community of conference attendees (mean: 4.17). The virtual background also enabled participants to increase their visibility at the conference (mean: 3.54) and to meet new people (mean: 3.54).
The idea of virtual backgrounds was warmly received by the interviewees; it helped in achieving a consistent graphic style for the conference. This increased the feeling that the participants were 'in one room'; that the various speeches and the speakers answering questions afterwards were part of a single event (\textit{'It's a very good idea; it gives the feeling that we're in one physical location'} [P4]).
This feature allowed participants to show themselves to others, acting as the digital equivalent of a business card ('\textit{It was linked to who this person is and where they come from. It is really helpful in being more engaged}' [P2]).

\begin{figure*}[h!]
  \includegraphics[width=0.9\textwidth]{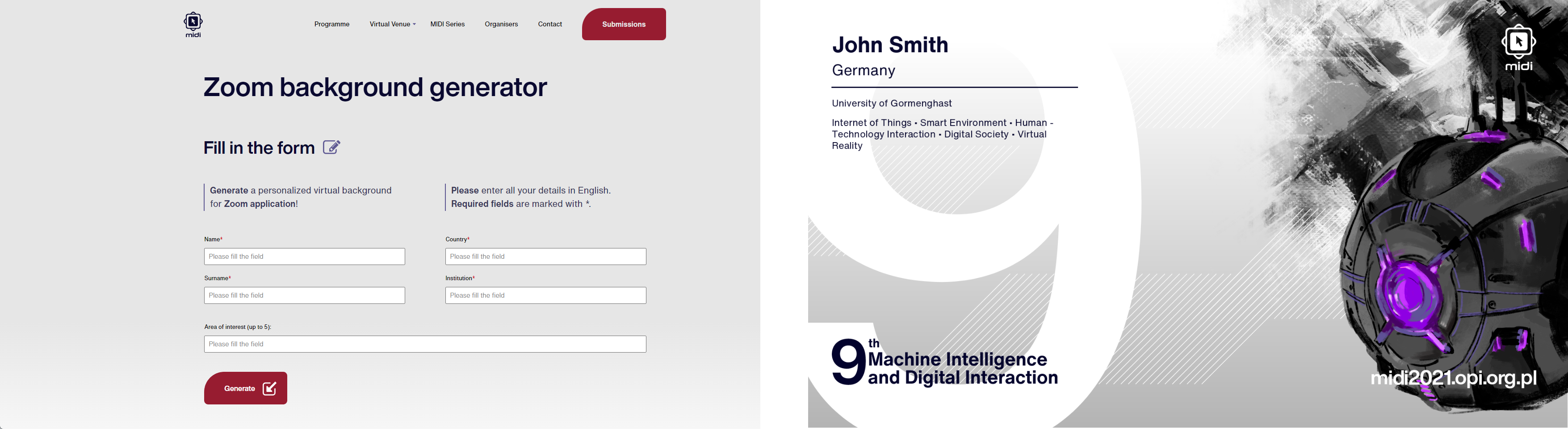}
  \caption{Illustration of the Virtual Background Functionality with the website generator for conference's participants (left) and the resulting background (right)}
  \Description{Figure 3 shows two screenshots. The first one, on the left, is presenting the interface of the conference's website page with a simple form for participant's name, surname, country, affiliation and research interest. The second one, on the right, presents generated background containing conference's name, logotype and data from the form.}
  \label{fig:illustration5}
\end{figure*}


It seems that the visual attractiveness of the conference's graphic design is imperative. 
At in-person scientific conferences, participants commonly take photographs of themselves in front of banners, roll-ups, or LED screens that depict conference logos and share them on social networks. This entails greater difficulty during remote conferences. One respondent stated: \textit{'I wonder, for example, whether somebody who has attended the conference has pictures of that event. I was thinking how online conference could have this}' [P1]. A graphically attractive virtual backdrop could also provide participants opportunities to take 'screen shots' of themselves and the background, and to share such images on social networks.
As all participants could use the virtual background (not only those who presented), this functionality instilled the feeling that every participant was a real part of the event ('\textit{It was easy. It was useful. It gives the feeling that you are a real participant}' [P5]). 

\subsection{\textbf{Map with Attendee Location Pins---73 locations pinned}}
In addition to enhancing attendees' sense of belonging to a larger community of conference participants (mean: 4.04), pinning information about themselves to a virtual map allowed them to increase their visibility at the conference and meet new people (mean: 3.38 and 3.46, respectively).
Participants received the pinning of the virtual map with enthusiasm. It acted as the equivalent of a badge. ('\textit{a very nice thing. It gives you a live picture of attendees. I think that each author should be encouraged to attach a pin}' [P1]). Comparisons were drawn between this functionality and the paper-based attendee lists of in-person conferences: '\textit{There used to be conference lists, but now, if there's personal information, there's no list of participants. Here it was very cool that whoever wanted to could insert themselves. And whoever inserted themselves had the equivalent of this list. It's a very good idea}' [P3].
One respondent was unsure of the purpose of such a functionality. This demonstrates that its use could be better explained during announcements and publicised on the conference website.

\begin{figure}[h!]
  \includegraphics[width=0.5\textwidth]{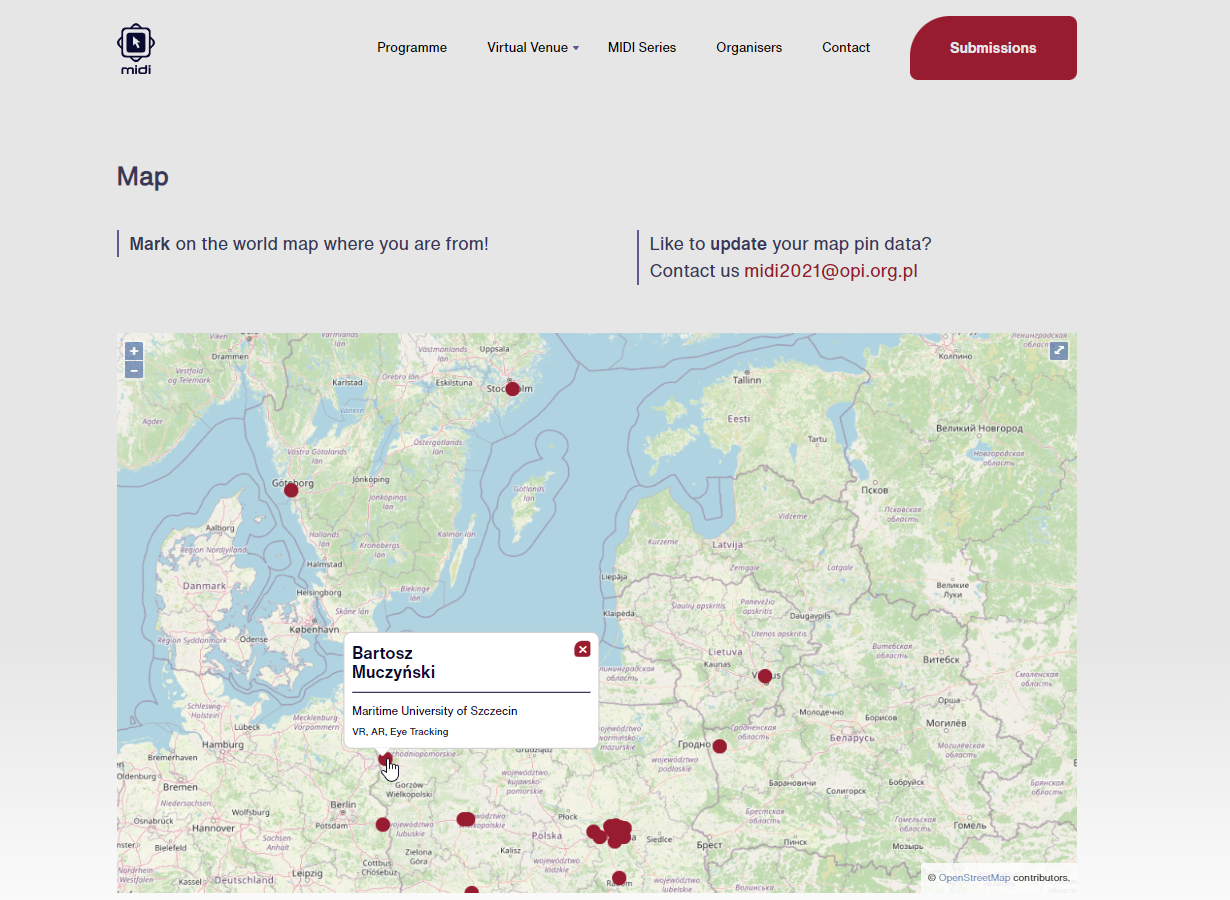}
  \caption{Illustration of the Location Pin Functionality}
  \Description{The figure 4 presents the screenshot from the conference's website with the interactive map where participants could pin their location. Pin with one of co-author's name, surname, city and research interest is shown.   }
  \label{fig:illustration1}
\end{figure}

The aim of this functionality was to increase attendees' visibility to one another. In contrast to a virtual background, this can be achieved using pins on a virtual map even when participants do not use cameras. 
At in-person conferences, attendees perceive the presence of others by sight and by sound. They can see how many other attendees occupy the same room. Wearing name badges around their necks enables attendees to discover information about each other without explicitly asking. All of this facilitates conversation. 
During remote conferences, we can see how many are attending a session if the platform's statistics allow it (this possibility existed at the conference in question). While participants do not always enter their real names and surnames, they are often visible under nicknames; otherwise, not much is known about them.

\subsection{\textbf{Virtual Mural---59 images added}}

Conference attendees perceived the virtual mural to hold primarily a socialising function. It increased participants' sense of belonging to the larger conference community (mean: 4.06) and fostered exchanges between hitherto unfamiliar people (mean: 3.24).
The attendees welcomed the opportunity to create a virtual wall using their own photographs. The concept was somewhat familiar to the attendees, as it is also customary following in-person conferences for photo galleries to be published on their websites ('\textit{Very nice idea; most of the conferences have picture galleries}' [P1]).
The interviewees also appreciated the novelty: ('\textit{And when we did our own conference, we asked people to turn on their cameras and we took a picture of the screen, with the participants...But everyone was standing in their own compartments. And here it was...such a cool idea that you could ask anyone to take a photo and that they were so spread out, these photos. And there was no feeling that everyone was sitting in their own little cage}' [P3]).
The virtual mural was firstly an attempt to document 'oneself at the conference'; secondly, to see who else was at the conference; and thirdly, to co-create. Breaking the monotony of conference receptions is an important function. At remote conferences---even if the activities vary (e.g. presentation sessions, answers, and announcements)---most of them occur in passive reception mode, i.e. the recipient is listening. The virtual mural provided an opportunity for active participation and to increase attendees' engagement ('\textit{It was very creative and people liked this idea. It really keeps you engaged and holds your attention---because nowadays, we are quickly distracted'} [P2]).
One of the five interviewees was unaware of the virtual mural and had failed to notice its link on the chat platform. This may indicate that participants were less likely to notice the features not publicised on the conference website, but announced solely by presenters during breaks between sessions.



\subsection{\textbf{Introductory Video with Tips---viewed 94 times}}
 In terms of meeting the goals of the conference, the introductory video\footnote{The video can be accessed via YouTube url: \url{https://www.youtube.com/watch?v=IbsX6z5DgQI}} increased participants' sense of belonging to a larger conference community (mean: 3.52). Of the five interviewees, only one had viewed the video. It was the least known and used of the six functionalities. The interviewee who watched the video found it useful: '\textit{the video helped me to understand what and how. It showed me exactly what to do}' [P4]). It seems that the low popularity of the video may have been influenced by its link's inclusion at the end of the email that contained attendees' login details sent the day before the conference. The video itself was not subsequently mentioned during the introduction or announcements.


\begin{figure}%
    \centering
    \subfloat[\centering]{{\includegraphics[width=0.4\textwidth]{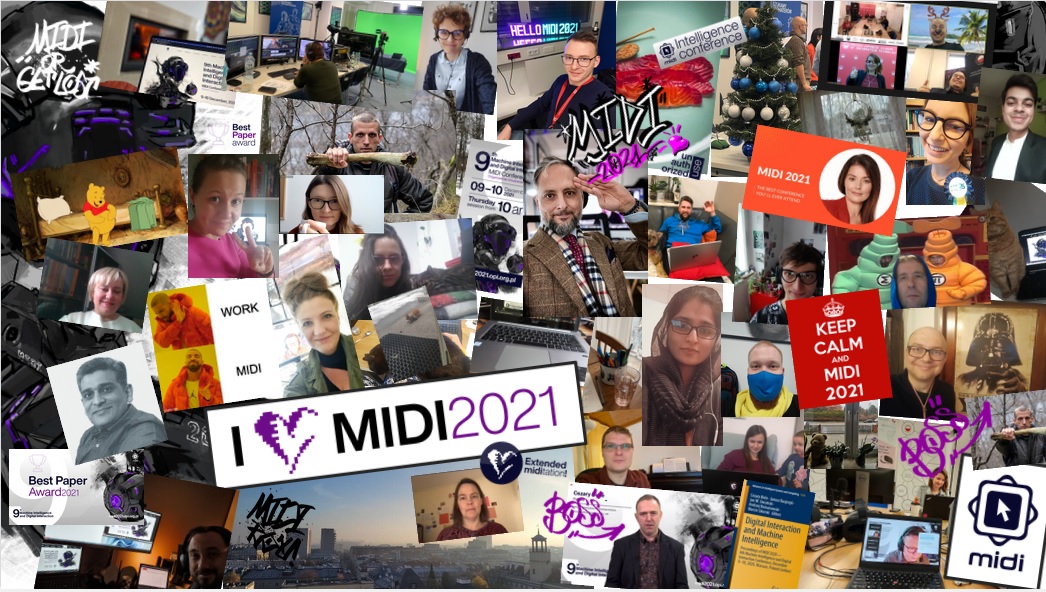} }}%
    \qquad
    \subfloat[\centering]{{\includegraphics[width=0.4\textwidth]{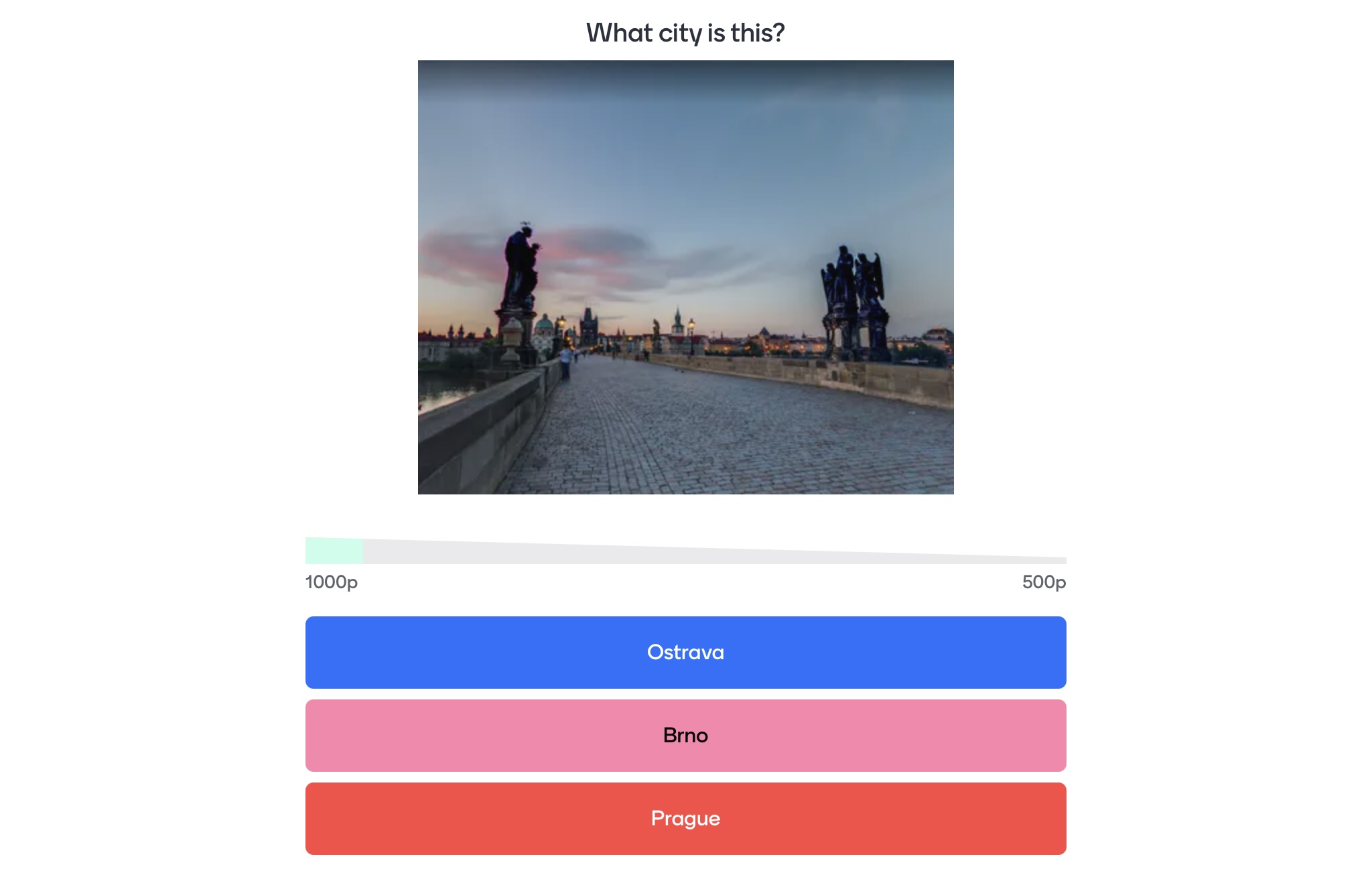} }}%
    \caption{Illustration of the virtual mural functionality (a) and a geo-game quiz as seen by conference's participants (b)}%
    \Description{Figure \ref{fig:muralgame} presents two figures. On the left there is so called virtual mural that was created by conference's participants from pictures taken by them and uploaded via provided link. It shows photos of people taken during the conference presenting how and where they attended it. Photos are spread randomly on the picture creating a form of a digital collage. On the right a quiz game panel is visible with a photo of a important landmark in some city and three options with different cities names. }
    \label{fig:muralgame}%
\end{figure}

\subsection{\textbf{Social Event (Geogame)---23 participants}}
The geogame also enabled attendees to feel that they were members of the conference community (mean: 3.53). Two of the five interviewees had participated in the game, which involved guessing in which cities a series of photographs had been taken. Participants indicated that it was an interesting experience that served to break the monotony of the day: '\textit{It was a good idea; it helped participants to keep their attention. I know that all attendees are able to participate in a game like that}' [P2]).
Information about the game appeared only during announcements of the chair; it was not publicised on the conference website, in the programme, or in mailings. This may be why many participants overlooked the game ('\textit{I missed something. Maybe there was a link in the chat; perhaps I didn't notice. I also had lectures on Friday, so I was a little disconnected from that. Then, maybe I missed something...The game missed me because I didn't find it}. [P3]').

\subsection{General Discussion}

\subsubsection{A different way to experience a remote conference}
In light of the interviews, it appears that slightly different modes of experience and senses of belonging can be observed between remote and in-person conferences. In an in-person environment, the participants engage in a physical space, such as a hotel or a conference hall. Sometimes, they change rooms to attend simultaneous tracks, or change seats within the same room. They are able to see the other participants. The experience of attending an in-person conference is deeply embodied. At a remote conference, participants typically remain in a room that is cognitively familiar, such as their own home, office, or workplace; an inferior sense of presence is experienced. 
Another difference lies in the 'entrances' to conferences themselves. In an in-person environment, the participants physically move their bodies from their hotel rooms to the conference venue, for example. This change of space accentuates the commencement of the conference experience. Even when distracted by their mobile phones, participants have the feeling that they are peering into other activities at the conference. At a remote conference, the opposite is true: the stream that is running often constitutes only one of the activities (professional, social etc.) that demands participants' attention. A remote conference is a place 'to look into'---sometimes one of many. This was well articulated by one interviewee: '\textit{Disadvantages---that everyday life pursues us. If I went away, maybe I would participate more: I could focus; I wouldn't have lectures; I could delegate. Maybe the personal contact would be better}' [P3]. This different mode of reception entails greater difficulty in the delivery of information to the participants. In-person conferences adopt a multi-channel approach: billboards, posters, programmes available in hard copy (usually distributed as part of a welcome packet), and badges all contribute. These elements are accompanied by announcements delivered by the chairs. Moreover, in-person conferences are social events: the participants are surrounded by others, and even if they lack knowledge of the programme, they can discover it from fellow participants---during a conversation over lunch or coffee, for instance. In other words, as well as programmes on hard copy and online, participants themselves serve to propagate information about the conferences.

\subsubsection{A different mode of communication with participants}
With the above in mind, conducting a remote conference requires a different approach. Everything must be explained in the programme. Organisers cannot expect participants to be informed solely by announcements or by other participants. In addition, illustrative visual material should be displayed during announcements. Presenters must repeatedly inform participants of additional activities and contextualise those activities: what purpose they serve, what benefits attendees might derive from them (e.g. becoming visible to others, facilitating interaction, establishing interesting contacts, or collaborating in research). Incorporating virtual bulletin boards---containing dynamically-created notes on what awaits attendees on a given day---is also worthy of consideration.

\subsubsection{Networking}
It might seem that in the age of social media and the ubiquitous presence of the internet, merely knowing somebody's name enables contact with that person; yet this is not always the case. There is a need at conferences for quickly contacting other attendees. Our conference incorporated a chat system for this purpose, which included a general channel and a direct messaging function. It transpired, however, that not all participants followed the chat. At an in-person conference, a direct message from another attendee cannot be ignored; this is by default not the case during a remote conference. On the Zoom platform, messages from individual participants can be lost when mixed with others that are addressed to everyone. Tellingly, of the five interviewees, only one was contacted via chat; the others failed to notice messages. The email channel transpired to be significantly more effective for this purpose, but it poses other difficulties. For data protection reasons, the organisers cannot distribute a mailing list, and participants frequently decline to consent to sharing their contact information during registration. This leads to a paradox in which attendees can be present at the same conference online (or even interacting in the same breakout room), but are unable to continue their discussions after the sessions have expired.
One solution involves the organisation of open breakout rooms in which participants can arrange to converse---even while other sessions continue. Another solution involves the introduction of a private chat functionality at the conference website, which would be accessible after logging in.

\section{\textbf{Recommendations and Future Work}}

Based on this research, we recommend that remote academic conferences focus on consistent multi-channel communication to facilitate attendees' acquisition of knowledge on the multimedia activities on offer. This includes e-mail communication, session-by-session schedule updates, and 'next up' television-style subtitles. Next, the provision of space for 'virtual murals' or functionalities that facilitate the collective community experience can be of great benefit. These might be used for sharing conference photographs or for disseminating knowledge on conference activities. We also recommend conference-themed photo frames, badges, and stickers to complement the common images shared on social media. We advocate the provision of introductory conference and technical training (e.g. in the form of pre-recorded videos that are mentioned during pre-conference communication and live sessions) to all participants on the ICT tools to be used, such as entering Zoom breakout rooms and using interactive whiteboards. Moreover, to encourage active participation, authors should be encouraged to actively gather feedback from their audiences and seek their opinions on matters of interest. Finally, we recommend that activities such as the placing of pins on virtual maps or the playing of games be conducted collectively during breaks in a workshop-like structure; this will enable more meaningful interaction between sessions. Aside from these findings, more research is needed both to further verify our insights and to explore opportunities for organising scientific events in the virtual space. Immersive online conference experiences are a particularly promising and necessary research direction, as such multimedia events are of key importance to the growing international scientific community - especially at this time when remote conferences are becoming the new normal.


\bibliographystyle{ACM-Reference-Format}
\bibliography{bibliography}










\end{document}